# Good Friends, Bad News
# Affect and Virality in Twitter

Lars Kai Hansen[1], Adam Arvidsson[2], Finn Årup Nielsen[1],
Elanor Colleoni[3] and Michael Etter[3]

[1] DTU Informatics, Technical University of Denmark,
DK-2800 Lyngby, Denmark, Email: lkh,fn@imm.dtu.dk

[2] University of Milan, via Conservatorio Milan, Italy,
Email: adam.arvidsson@unimi.it

[3] Copenhagen Business School, DK-2000 Frederiksberg,
Email: elc.ikl,me.ikl@cbs.dk

January 4, 2011

**Abstract**

The link between affect, defined as the capacity for sentimental arousal on the part of a message, and virality, defined as the probability that it be sent along, is of significant theoretical and practical importance, e.g. for viral marketing. A quantitative study of emailing of articles from the NY Times (Berger and Milkman, 2010) finds a strong link between positive affect and virality, and, based on psychological theories it is concluded that this relation is universally valid. The conclusion appears to be in contrast with classic theory of diffusion in news media (Galtung and Ruge, 1965) emphasizing negative affect as promoting propagation. In this paper we explore the apparent paradox in a quantitative analysis of information diffusion on Twitter. Twitter is interesting in this context as it has been shown to present both the characteristics social and news media (Kwak et al., 2010). The basic measure of virality in Twitter is the probability of retweet. Twitter is different from email in that retweeting does not depend on pre-existing social relations, but often occur among strangers, thus in this respect Twitter may be more similar to traditional news media. We therefore hypothesize that negative *news content* is more likely to be retweeted, while for non-news tweets positive sentiments support virality. To test the hypothesis we analyze three corpora: A complete sample of tweets about the COP15 climate summit, a random sample of tweets, and a general text corpus including news. The latter allows us to train a classifier that can distinguish tweets that carry news and non-news information. We present evidence that negative sentiment enhances virality in the news segment, but not in the non-news segment. We conclude that the relation between affect and virality is more complex than expected based on the findings of Berger and Milkman (2010), in short 'if you want to be cited: Sweet talk your friends or serve bad news to the public'.

# 1 Introduction

Viral communication is already of significant practical importance and the scientific interest is increasing. The scientific interest derives in part from the light shed by viral communication on meme diffusion and opinion formation. In the growing viral communication literature there has been an interest in affect and emotion as crucial factors behind successful viral diffusion. The idea has been that people are more likely to send on material that like 'pets, sex and the absurd' is able to actualize a common experience of affective arousal, be this laughter, compassion, anger or surprise. This idea that affectively charged viral messages are more likely to spread than affectively neutral ones has been present within the more anecdotic marketing literature [1, 9], as well as in the more systematic aggregations of qualitative studies that have arisen in recent years. For example, in their study of word of mouth (WOM) marketing, Kozinets et al. [14] argue that such messages are more likely to be taken seriously and further diffused by consumers once they are inserted within a network of affectively significant relations that is able to transform messages from *'persuasion oriented 'hype' to relevant, useful, communally desirable social information that builds individual reputations and group relationships'*. It is in this *'transformation of a market narrative into a social one'* that *'the WOM communicator performs* [the] *services* [that] *are valuable to the marketers'*[14]. Similarly, based on a qualitative analysis of nine viral marketing campaigns, Dodele et al. [5] conclude that the key to success is the ability to stir up an emotional arousal among the people who pass along the message. Indeed, as Vincent Miller [20] argues, communication on social media can be primarily understood as 'phatic', that is, geared towards the creation and consolidation of affectively significant relations, rather than towards the transfer of information.

Only recently however has the hypothesis of the effect of affective charge on viral diffusion been tested in a large-scale quantitative study. In their recent work, Berger and Milkman [2] use a sample of 6,956 articles from the New York Times articles published between August 30th and Nov 30th 2008. The authors conclude that there is a strong link between affect, as measured by a sentiment analysis of article content, and whether content is highly shared; as defined as whether the particular article made the New York Times' list of 'most emailed' articles. They also conclude that positive content is more viral than negative content, but that articles with some negative content, like anger or anxiety are both more likely to make the paper's most emailed list. These results hold controlling for how surprising, interesting, or practically useful content is, as well as external drivers of attention, like how prominently articles were featured. Berger and Milkman's suggestion that affective charge has a discernable impact on viral diffusion is corroborated by a host of quantitative studies. It is also supported by common sense. In a social media environment where social relations have effectively become a medium of communication, content that is more likely to activate such relations is also more likely to spread.

Interestingly a similar relationship has been suggested, if implicitly, by a long range of sociological research on the diffusion of ideas and innovations, from the work of Tarde [32] via that of Lazarsfeldt [16] and Rogers [25] without the terms 'virality' or 'affect' ever being used. However, Berger and Milkman also claim that their conclusions about the link between positive affect and viral diffusion has a general validity, and they found this link on psychological theories, like the claim that '*consumers often share things to self-enhance [...] or communicate identity, and consequently positive things may be shared more because they reflect positively on the self*'[2]. In other words, they suggest that it is part of human nature to share positive content more often than negative content. However, their conclusion appears to be in contrast with classic theory of selection and diffusion in news media [8], which emphasizes negative affect as promoting propagation. This applies in particular to the theory of news factors. News factors are '*relevance indicators that not only serve as selection criteria in journalism but also guide information processing by audience*' [6]. News factors research can be traced back to 1922, when Lippman [18] introduced the first model of news values. In part inspired by cognitive psychology it has developed into several different models, see e.g., [8, 29, 26] that aim to explain why certain events obtain more media attention or readers' awareness then others. The notion of news factor *negativity* has been introduced by Galtung and Ruge in 1965 [8] and since then has been taken into account by other models [6, 26, 29] partly splitting up negativity in subcategories as for example *conflict*, *damage*, *aggression* or *failure*. Several studies show that the news factor negativity has a significant impact on readers' awareness as well as on journalists' selection [10, 22, 28, 30]. Straughan's study [30] shows that news items containing conflict had a significant positive impact on reader interest in news stories. Similar, Schulz [27] found that events with the news factor aggression obtained high awareness. Further, the role of negativity in news selection has been highlighted by Peterson [22], showing that conflictual events are rather published than cooperative events. In a comparison of different news factors Harcup and O'Neill [10] range negativity among the highest news factors. The most recent study on news selection by Schwarz [28] provides evidence that the news factor negativity correlates significantly with assigned text space in news media. The contrast between Berger and Milkman's [2] findings and established research on news factors can be resolved by distinguishing between different communication media. We suggest that while Berger and Milkman's findings might apply to email networks that are marked by strong reciprocity, not all networks of information diffusion in social media are reciprocal. Such platforms may support diffusion dynamics that are more similar to those of conventional news media.

## 1.1 Twitter

In recent years, microblogging, where users can describe their current status in short posts, has emerged as a new important communication platform [13]. The most popular microblogging platform is Twitter. Twitter connects both friends like in the symmetric Facebook network and users with shared interest like one-way structures like news groups or conventional blogs. In-links in Twitter are called followers, while out-links are called friends or 'followings'.

Business entrepeneurs Naval Ravikant and Adam Rifkin [23] suggest that Twitter's value is increased by the fact that Twitter is in part an interest graph, thus revealing more of the users behaviors than a purely friends-based social graph. Their notion of an interest graph describes a network that differs from a social graph in four important respects: One-way following rather than two-way reciprocal relationship; it is organized around shared interests, not personal relationships; it is public by default, not private by default; it is 'aspirational': not who you were in the past or even who you are, but who you want to be [23].

Without making reference to Ravikant and Rifkin, Weng et al. challenges the notion of Twitter as an interest graph. Based on a quantitative study they reveal many similarities between Twitter and conventional social networks [33]. Weng et al. study a corpus of tweets created by the most followed twitterers in Singapore. The total number of tweets collected from these accounts was $1,021,039$, with the majority dated between April 2008 to April 2009. Weng et al. find that a large fraction ($>74\%$) of their population share symmetric links with the majority of their friends ($>80\%$), furthermore they find notable assortative mixing also referred to as homophily, i.e., the tendency to follow twitterers with the same number of followers as yourself, and also a powerlaw degree distributions; both are characteristics of social networks. Yet, in a topic model of the tweet texts, Weng et al. report evidence that Twitter friends share interests (topics), hence, connect with the idea that Twitter relations reveal similar interest and behaviors.

Thus while Twitter users might entertain reciprocal relations with some followers, information diffusion through retweets tends to expand far beyond this circle of reciprocity, and proliferate through the 'interest graph', among followers with whom no reciprocal relations are entertained [12]. A large scale study by Kwak et al. [15] supports the notion that Twitter is not simply social network. They crawled the entire 'Twitter sphere' to find no sign of a global power law link distribution and they report low reciprocity in contrast to the results of the more geographically localized study of Weng et al. [33].

*Retweeting*, is the mechanism by which a user quotes another user and is one of the important means of meme propagation and opinion formation in Twitter. Meme propagation was studied recently in context of blogs and news media by Lescovec al. [17]. In contrast to the use of quotes (".."') used to track memes in the work of Lescovec et al., in Twitter it is explicitly indicated which user is cited by use of either the 'RT @user' or 'via @user' notations.

Retweet practice is a topic of significant current interest. Honeycutt and Herring [11] and Boyd et al. [3] both discuss retweet syntax, and the role of retweeting in Twitter discourse. Boyd et al. note three mechanisms supporting meme propagation, including the use of topical tags '#topic', mentions of given user '@user', and finally the use of shortened URL's to allow receivers to access background sources of a given meme. Boyd et al. explicitly asked a Twitter community what and why they retweeted. The feedback revealed a quite complex set of retweeting mechanisms, including (self-)branding, general news interest, and encouraging social activity amongst many other reasons.

A more quantitative study of meme propagation in Twitter is reported in Suh et al. who build simple models of the probability of being retweeted [31]. Suh et al. collected $10,000$ tweets and subsequently found that 291 of these had been retweeted (mere 3%). They build a model of retweet probability based on context variables mentioned above and further includes basic aspects of the graph structure, i.e., the number of followers and friends of the tweeter. The study confirms that inclusion of URL and hash tags both improve the probability of retweet, while explicit mentions of another user seem to reduce retweet probability, although this finding is only a trend in the data ($p \sim 0.07$).

Thus Twitter is an interesting and complex communication platform serving both as a social network and as a new medium of information sharing. Thus when asking 'what are the determining factors for my message to be retweeted' it may depend on both on the type of content and whether the communication is intended for a broader audience or for a more closed community of friends. In case of addressing a broader audience with news content the message sentiment may be an important determinant. Following this analysis we set out answer the following research questions

- Q1 How accurately can text be characterized as 'news'?
- Q2 How big a fraction of Twitter is news?
- Q3 If Twitter is a news medium, does negative sentiment influence virality?
- Q4 Does sentiment influence retweet probability diffentially in news and social messages?

The paper is organized as follows: In the next section we define the methodology, including Twitter samples and the statistical models. In the following section we present the results regarding news classification and sentiment detection, and in a final section we discuss the findings, provide conclusions and ideas for further study.

## 2 Methods

To address the research questions Q1-Q2 we will use a simple machine learning method - a Naive Bayes classifier - to detect whether a message is 'news'. This classifier is trained and tested from a labeled text corpus, the Brown Corpus. The trained classifier is applied to two medium large Twitter samples containing a 'complete sample' of a discourse driven by a news event and a sample of randomly selected tweets, respectively. In both Twitter samples we estimate sentiments and we build generalized linear models to investigate whether sentiment influences virality in terms of the retweet probability.

### 2.1 Text and Twitter corpora

We use three different text corpora to test the above questions,

C1 BROWN, a general text corpus with a known mixture of news / non-news documents [7]. The corpus consist of $N_B = 47,134$ sentences of which $4,623$ are categorized as news.

C2 COP15 a Twitter data set that is designed to comprise a complete set of tweets for a specific news driven vent. COP15 refers to the The 2009 United Nations Climate Change Conference that took place in Copenhagen, Denmark, between December 7. and December 18. The conference included the 15th Conference of the Parties (COP 15) to the United Nations Framework Convention on Climate Change[1]. The conference and the international activism that took place during the conference were extensively covered in news media and in the blogosphere. A total of $N_C = 207,782$ tweets were downloaded during the month of December 2009 by querying the Twitter Search API[2] with the term 'cop15'.

C3 RANDOM a random sample of tweets. $N_R = 348,862$ tweets were downloaded with the Twitter streaming API[3] during the time interval from September 9. to September 14., 2010. The Tweets were randomly sampled following the 'Spritzer' protocol[4].

## 2.2 Language analysis

The sentiment analysis is designed to work on English tweets. Downloaded tweets are provided with an indication of the language. However, for many non-English tweets the field is set to its default value 'English'. We therefore first constructed a language filter: After download a language detector scored each tweet for *englishness*. The language detector used a list of words manually scored from $-3$ to $+3$ indicating an estimate on how English each word is. The englishness of the words in a tweet was accumulated, and tweets with a positive sum were regarded as English. For the streaming Twitter data 106,719 remained after extracting tweets where the user had set the language to English and the language detector also detected the tweet as English. The language filter eliminated only a few percent of the COP15 tweets, while more than 60% of the RANDOM data was removed as non-English, see table 1 for details.

## 2.3 Naive Bayes news classifier

To determine the 'newsness' of a tweet we trained a classifier on the sentences of the Brown corpus[5] with the NLTK toolkit [19]. The news category of the Brown corpus has $4,623$ sentences, while other categories have a total of $42,511$ sentences (we excluded the 'editorial' category). The classifier is trained on a bag-of-terms representation. A stop word list on 571 words excluded common words and the 10,000 most frequent words were extracted from the corpus and used as terms. For the RANDOM data 'lol', 'love' and 'good' were the most frequent terms, while for the COP15 data the most frequent terms were 'obama', 'world', and 'deal', reflecting the latter is driven by a news event. A sentence is represented in the classifier as $D = 10,000$-dimensional vector $\boldsymbol{w}$ where the $d$'th entry, $w_d$ is either 1 or 0 depending on whether the $d$'th term is present or not in the given sentence.

---
[1]http://www.denmark.dk/en/menu/Climate-Energy/COP15-Copenhagen-2009/cop15.htm
[2]http://dev.twitter.com/doc/get/search
[3]http://apiwiki.twitter.com
[4]http://dev.twitter.com/pages/streaming_api_concepts
[5]http://icame.uib.no/brown/bcm.html

We use the so-called Naive Bayes classifier based on univariate discrete distributions [24], i.e., we posit a model of the probability of a tweet being news of the form

$$\begin{aligned} p(\text{news}|\boldsymbol{w}) &= \frac{p(\text{news})p(\boldsymbol{w}|\text{news})}{p(\text{news})p(\boldsymbol{w}|\text{news}) + p(\neg\text{news})p(\boldsymbol{w}|\neg\text{news})} \\ &= \left(1 + \frac{p(\neg\text{news})\prod_{d=1}^{D} p(w_d|\neg\text{news})}{p(\text{news})\prod_{d=1}^{D} p(w_d|\text{news})}\right)^{-1} \end{aligned} \quad (1)$$

where $p(w_d|\text{news})$ is the probability of occurrence of the $d$'th term in news category of all training tweets. All terms are present in at least one news and one non-news tweet.

Splitting the corpus in 75% for training and the rest for testing, the NLTK naive Bayes classifier will report the test error which is an unbiased measure of performance.

We also apply the trained classifier on the tweet data getting a probability of 'newsness' for each tweet in RANDOM and COP15. Based on the set of probabilities for all tweets we calculate the rate of news tweets, i.e., fraction of tweets with $p(\text{news}|\boldsymbol{w}) > 0.5$.

## 2.4 Sentiment scoring

For English tweets sentiment was estimated via an English word list manually curated for Twitter. Thus, we follow the classical approach used for sentiment scoring in conventional English text [4], but with a dedicated wordlist. The present list associates $1,446$ words with a valence between $-5$ and $+5$. Example words from the top of the listed below:

```
abandon       -2
abandons      -2
abandoned     -2
absentee      -1
absentees     -1
aboard        1
abducted      -2
abduction     -2
```

Sentiment estimation results in a *valence* and an *arousal* score for each tweet. The valence of a tweet $n$, $v_n$, is computed as the sum of the valences of the individual words in the tweet $v_{n,i}$, while the arousal $a_n$ the sum of the absolute value of the valences,

$$v_n = \sum_i v_{n,i} \qquad \text{valence} \qquad (2)$$
$$a_n = \sum_i |v_{n,i}| \qquad \text{arousal.} \qquad (3)$$

## 2.5 Modeling the retweet probability

The extracted features were included in a generalized linear model (GLM) assuming a binomial distribution with the standard link function [21]. Denoting the probability of retweet by $p(RT|\boldsymbol{f})$, where $\boldsymbol{f}$ is a set of $F$ features derived from the tweet, the GLM reads,

$$p(RT|\boldsymbol{f}) = \left(1 + e^{-\sum_{i=0}^{F} \beta_i f_i}\right)^{-1}. \qquad (4)$$

The coefficient $\beta_0$ of the 0'th feature $f_0 = 1$ ensures proper normalization. The coefficients $\beta_i$ are estimated using iterative likelihood maximization. The difference in log-likelihood of two nested models is approximately $\chi^2$ distributed for large data, hence, can be used to test hypotheses about the relevance of individual features. In particular we estimate a model with all features and $F$ sub-models with a single feature removed. The $t$-statistic values in table 1 express the relevance of the given feature to the retweet model.

We formulate the modeling problem as in [31]: Imagine you are a tweeter who wants to be retweeted, how should you formulate your tweet? However, relative to [31] we introduce three modifications. First, we focus on features that actually can be manipulated at the time of tweeting, this includes the presence of '#', '@', and 'URL', but not the variables related to the graph structure, since these variables can not be manipulated at time of tweeting. The second modification is that we flag a tweet as a retweet if the text has a pattern with 'RT' or 'via' followed by a user name ('@user'). In [31] the authors located retweets of an initial pool of tweets. Our approach leads to a larger sample size of retweets, but may suffer from an unknown bias related to text modifications introduced by the retweeting party. We expect this bias to be limited and for retweets based on Twitter's 'retweet button', there is no bias. Our criterion also includes possible 'retweets of retweets'. Finally, we include the presence of negative sentiment as a covariate, hypothesizing that this may help explain propagation as is hypothesized in news media. User names that are part of the retweet indicator are not counted as a separate mentioning ('@').

To further test the role of newsiness on we create a feature which is the logical combination of news and negative.

Table 1: Estimated t-values within the RANDOM and COP15 data for the binary variable encoding the presence of negative sentiment. The t-value is obtained in a general linear model (logistic regression) of the retweet probability versus the independent binary (present/not present) co-variates: negative sentiment, hashtag, mention, and url.

|  | | COP15 | | Stream | |
| --- | --- | --- | --- | --- | --- |
|  | | All | English | All | English |
| N | | 147,041 | 136,262 | 335,236 | 106,719 |
| Rate of News | | 0.303 | 0.305 | 0.226 | 0.233 |
| t(Negative) | — | 4.889 | 4.649 | 2.775 | -0.024 |
| t(Negative×newsness) | — | 2.275 | 1.471 | -6.019 | 3.904 |
| Only tweets with Arousal > 0 | | | | | |
| N | | 44,611 | 42,087 | 53,473 | 51,929 |
| t(Negative) | — | 3.276 | 2.372 | -9.725 | -9.374 |
| t(Negative×newsness) | — | 1.125 | 0.180 | 1.179 | 1.239 |

# 3 Results

The results are discussed in relation to the four research questions Q1-Q4. Details are summarized in table 1.

### Q1: The Naive Bayes classifier can detect news

We assign 75% of the sentences in the Brown corpus for training leaving the 11783 for testing. The NLTK naive Bayes classifier reports an accuracy of $84\% \pm 1\%$ on the test data.

### Q2: 23% of all tweets are news

We apply the classifier to the tweets in the two sets RANDOM and COP15. In the RANDOM sample 23% of the tweets meet the criterion of having a probability of news larger 0.5. If biased, this could be a slight underestimate, as the fraction of news items (i.e., the a priori class probability) in the Brown corpus is about 0.1. In the COP15 sample we find that a larger number of tweets, 31%, are detected as news.

## Q3: Negative sentiment does not promote retweeting in the RANDOM sample

We trained a number of generalized linear models to test for the significance of negative sentiment in the COP15 and the RANDOM samples. In the RANDOM sample the weight for presence of negative sentiment is slightly negative, however the t-statistic indicates that we can not reject the null that there is no effect of negative sentiment. However in the COP15 sample negative sentiment promotes retweeting and the positive weight is highly significant. If the analysis is restricted to the subset of tweets that have non-zero arousal, i.e., positive or negative content, the tendency found in the RANDOM sample of all tweets is strongly amplified, it is a strong promoter of retweet if content is positive.

## Q4: Negative sentiment does promote retweeting in news tweets in both the RANDOM and the COP15 samples

To test for the interaction between newsiness and negative sentiment we create a new multiplicative variable (negative sentiment)× (probability of news). This interaction variable promotes retweeting in both COP15 and RANDOM. However, if we confine ourselves to the aroused tweets the conclusion persists even if it was strongly favored to have positive content in the sample at large.

# 4 Discussion

There is an ongoing discussion on the graph structure and dynamics in the Twitterverse. A large scale quantitative analysis led Kwak et al. [15] to conclude that Twitter can be viewed both as a social network and a news medium, while the more localized study of Weng et al. [33] found structures more reminiscent of a social network. At the same time the work of Berger and Milkman [2] and general psychological arguments favor the sharing of non-negative content among friends, while classical theories of news diffusion points to increased attention to news if content is negatively framed. This seemingly paradoxical set of results are reconciled in our findings. Using a trained classifier that can reliably detect news, we find that about a quarter of the tweets in a random sample are news, while the more focused sample of tweets relating to the global news event COP15 has a higher fraction nearing one third of news tweets. We note that these figures for news content are higher than estimated in a small informal investigation by San Antonio based analytics firm Pear Analytics[6]. We are in process of developing additional ways of estimating and validating newsness. The differences in news content are reflected in how much we can accelerate retweet by negative framing of content. In a random sample of tweets there is a slight tendency that negative content hinders retweet, albeit it is only a trend, while in the more news driven COP15 sample, negative content is a strong promoter of retweeting. If we look at the interaction term, we find that negative news is more retweeted than positive news. For the non-news segment which may be dominated by social tweets our results support the idea that positive content increase the probability of diffusion [2], whereas in the news segment, our findings confirm the impact of negativity on news awareness and selection as proposed by classic news diffusion theory [6, 8, 26, 29].

---

[6]http://www.pearanalytics.com/blog/wp-content/uploads/2010/05/Twitter-Study-August-2009.pdf

# 5  Conclusion

We found that a simple Naive Bayes classifier can quite reliably detect the presence of news in short communication. We found a higher amount of news content than has previously been reported: 23% in random tweets and 33% in the COP15 data. Investigating the probability of retweeting we found in a generic setting that negative sentiment is detriment to retweeting, while news related content propagates better if negative. *Hence, if you want to be cited: Sweet talk your friends or serve bad news to the public!*

# Acknowledgement

We thank the Danish Strategic Research Councils for their generous support to the interdisciplinary project 'Responsible Business in the Blogosphere'.